\documentclass[twocolumn,showpacs,showkeys,prb]{revtex4}% Physical Review B
\usepackage{amsmath}
\usepackage{amssymb}
\usepackage{bm}
\usepackage{epsfig}
\usepackage{graphicx}
\usepackage{color}

\begin{document}

\title{Low-frequency spectroscopy of superconducting photonic crystals}

\author{A.N.~Poddubny$^{a}$\footnote{{\it E-mail address:}
poddubny@coherent.ioffe.ru}, E.L.~Ivchenko$^a$, Yu.E.~Lozovik$^b$}
\affiliation{$^{a}$~A.F.~Ioffe Physico-Technical Institute, 194021
St.~Petersburg, Russia\\ $^b$~Institute of Spectroscopy, 142190
Troitsk, Moscow region, Russia}

\begin{abstract}
Transmission, reflection and absorption of electromagnetic radiation and 
photon  dispersion law for 2D photonic crystals with superconducting 
elements are studied. The calculation of optical properties of photonic 
crystals is studied by layer-by-layer Korringa-Kohn-Rostoker techniques. The 
results of numerical calculations performed for an array of superconducting 
cylinders have been understood in terms of a simple analytical model. The 
controlling of optical properties of superconducting photonic crystal by 
temperature and magnetic field is discussed. The variation of 
superconducting component density with temperature leads, particularly, to 
significant reduction of the transmission peaks and even to a nonmonotonous 
behavior of some absorbance peaks.
\end{abstract}

 \maketitle

%\mbox{}\\ {\it PACS:} 42.70.Qs, 74.25.Gz, 74.78.-w\\

\section{Introduction}
Photonic crystals (PCs) consisting of metallic wires are a subject of 
growing interest during last years and are widely studied both theoretically 
[\onlinecite{efros,pendry,sakoda,wang,krokhin}] 
and experimentally [\onlinecite{pimenov,ricci}]. At 
low-frequencies, e.g., in the giga- and terahertz spectral regions, the real 
part of the dielectric function of metallic component is large and negative 
which allows the formation of the complete two-dimensional photonic band gap 
in these spectral regions. Moreover, the electromagnetic waves cannot 
propagate in such a system at frequencies lower than the so-called cutoff 
frequency $\omega _{c}$ [\onlinecite{pendry}]. If lattice constant of 
these structures lies in the millimeter and submillimeter ranges their 
fabrication is relatively simple. However, the damping of electromagnetic 
waves in metals can suppress many potentially useful properties of metallic 
PCs. The possible solution of this problem is the replacement of the PC 
metallic component by a superconducting (SC) component. The dielectric 
function depends on superconducting gap $\Delta $ (see, e.g., Refs. [\onlinecite{lozovik,dobryakov}]  and 
references therein) and SC state can be varied by alteration of external 
parameters, such as temperature and external magnetic field which provide a 
method for a control of the optical properties of metallic PCs. The 
dielectric losses are reduced essentially below superconducting transition 
in SC state, and recent experiments have shown that photonic band edges 
become sharper in SC metals [\onlinecite{ricci}].

The band structure of a superconducting PC was analyzed in Ref.~[\onlinecite{lozovik}]. Here we calculate the transmission, reflection, and 
absorption spectra and analyze controlling of these spectra by temperature.
%--------------------------------------------------------------------
\section{Problem definition and method of calculation}
\begin{figure}[b]
	\begin{center}
  \includegraphics[width=0.2\textwidth]{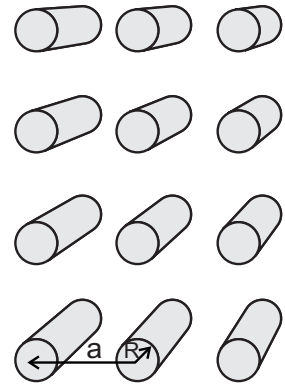}\\
  \end{center}
  \caption{Schematic representation of a two-dimensional photonic
  crystal with cylinders of the radius $R$ arranged in a square
  lattice with the lattice constant $a$. }
\end{figure}
We study the photonic crystal formed by a periodic array of parallel 
superconducting cylinders with radius $R$ arranged in a square lattice with the 
lattice constant $a$, see Fig.~1. Electromagnetic waves propagating in this 
structure are described by Maxwell equation 
\begin{equation}
\label{eq1}
\nabla \times \nabla \times \bm E(\bm r)=\left( {\frac{\omega }{c}} 
\right)^2\varepsilon (\bm r,\omega )\bm E(\bm r)
\end{equation}
where $\varepsilon (\bm r,\omega )$ is the dielectric 
function equal to that of the superconducting metal inside the cylinders and 
to unity outside them (in vacuum). For simplicity, we consider the 
electro-magnetic response of superconductor cylinders by using the 
two-liquid Gorter-Kazimir model and describe the electronic system as an 
admixture of two independent carrier liquids, superconducting and normal. 
With the increasing of temperature $T$ or magnetic field \textbf{\textit{B}}, 
the density of the normal component, $n_{n}$, increases at the expense of the 
superfluid component, $n_{s}$, because the total density of electrons 
$n_{tot}=n_{n}+n_{s}$ is conserved. At the critical temperature $T=T_{c}$ only 
normal phase survives whereas, at zero temperature, only superfluid phase 
does exist.

In the Gorter-Kazimir model the dielectric function of a superconductor can 
be presented in the following simple form
\begin{equation}
\label{eq2}
\varepsilon _{\rm GK} (\omega )=1-\omega_{p}^2 \left( {\frac{\alpha }{\omega 
^2}+\frac{1-\alpha }{\omega (\omega +i\gamma )}} \right)
\end{equation}
where $\omega _{p}$ is the plasma frequency $(4\pi n_{tot}e^{2}/m)^{1/2}$ corresponding to 
the total electron density $n_{tot}$, $m$ is the electron effective mass, 
\textit{$\gamma $} is the phenomenological damping coefficient describing the relaxation of 
normal electron component, and \textit{$\alpha $} is the fraction of superconducting component 
$n_{s}/n_{tot}$ which varies continuously with temperature (at magnetic field 
$H=$0) from $\alpha =1$ at $t=0$ to  $\alpha =0$ at $T=T_{c }$. It can be controlled also by magnetic field. It 
should be noted, that this equation is valid qualitatively for the 
frequencies \textit{$\omega $} below the superconducting gap 2$\Delta $.

At the critical temperature $T_{c}$ or critical magnetic field $H_{cr}$ the 
superconducting phase vanishes and the dielectric function can be described 
by the conventional Drude equation 
\[
\varepsilon _{\rm GK} (\omega )=1-\frac{\omega _p^2 }{\omega (\omega +i\gamma 
)}
\]
In the opposite limit, $T\to 0$, the contribution of the normal component is 
negligible and the dielectric function reduces to 
\[
\varepsilon _{\rm GK} (\omega )=1-\frac{\omega _p^2 }{\omega ^2}
\]
In numerical calculations we consider high-temperature superconductor 
YBaCuO; its parameters in the normal state are as follows: $\omega 
_{p}$=1.67$\times $10$^{15}$rad/s, $\gamma $=1.34$\times $10$^{13}$rad/s. 
For YBaCuO the superconducting gap can be estimated as $2\Delta \approx 6k_B 
T_c /\hbar $ ($k_{B}$ is the Boltzmann constant). For $T_{c}$=91 K one has 
\textit{ž$\Delta $}=28 meV which is equivalent to 7 THz. A value of 150 $\mu $m is chosen for 
the lattice constant $a$ so that frequencies of the first few allowed photonic 
bands lie below $\Delta $. The filling factor \textit{f=$\pi $(R/a)}$^{2}$ in our calculations is 
equal to 5{\%}.

The optical spectra are calculated for finite-size two-dimensional PC slab 
with the normal parallel to the symmetry direction $\left\langle {01} 
\right\rangle $. In this case the structure can be considered as $N$ identical 
layers, each containing one-dimensional chain of cylinders. In what follows 
we restrict ourselves to the case of normal incidence of TE-polarized light, 
which means that the electric field is parallel to the cylinder axis 
$z$, $\bm E=(0,0,E_{z})$ In this particular geometry Eq. (\ref{eq1}) can be 
rewritten as 
\begin{equation}
\label{eq3}
\left[ {\frac{\partial ^2}{\partial x^2}+\frac{\partial ^2}{\partial 
y^2}+\left( {\frac{\omega }{c}} \right)^2\varepsilon (x,y;\omega )} 
\right]E_z (x,y)=0
\end{equation}
The numerical calculation is performed by a two-dimensional bulk and 
layer-by-layer Korringa-Kohn-Rostoker techniques [\onlinecite{ohtaka}]. 

%------------------------------------------------------------------
\section{Results and discussion}
\begin{figure}[t]
  \includegraphics[width=0.48\textwidth]{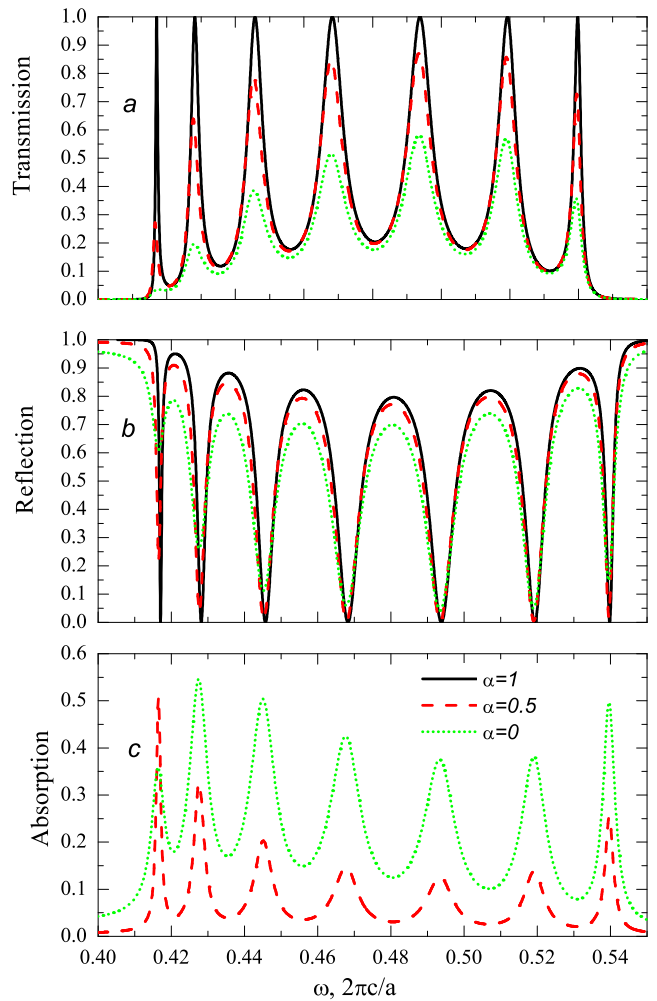}\\
  \caption{Transmission (a), reflection (b), and absorption (c) spectra of
a superconducting photonic crystal calculated for three different
values of $\alpha$ and the set of other parameters given in the
text. At $\alpha = 1$ the absorption is zero and the corresponding
line $A(\omega) \equiv 0$ coincides with the abscissa.}
\end{figure}
Figure~2 presents the transmission, reflection and absorption
spectra, respectively, $T(\omega)$, $R(\omega)$ and $A(\omega)=1 -
R(\omega) - T(\omega)$, for a $8$-layer-thick superconducting PC
calculated for different values of parameter $\alpha$. The chosen
spectral region covers the first allowed band, diffraction
channels at these frequencies are closed. For the pure SC phase,
$\alpha = 1$, the absorption vanishes, $R(\omega) = 1 -
T(\omega)$, and the transmission peak values reach unity. With
decreasing $\alpha$ the transmission and reflection at any fixed
$\omega$ decrease monotonously, with the transmission being more
sensitive to the variation of $\alpha$. At the same time the
absorption maximum can exhibit a non-monotonous behavior, at least
it is the case for the first absorption peak. For the chosen
lattice constant the parameter $2 \pi c/a$ is comparable with the
attenuation $\gamma$, the role of the electronic relaxation is
significant except for very low fraction of normal electrons, i.e.
except for the region $1 - \alpha \ll 1$. The oscillations in
optical spectra can be interpreted in terms of the interference
effects taking into account the modification of the photonic
dispersion in the PCs. Such spectral behavior is typical for
metallic PCs, see, e.g., [\onlinecite{wang}]. However, for the THz
frequency region considered here, the values of $|\varepsilon_{\rm
GK}(\omega)|$ are of the order of $10^4$ which is by almost three
orders of magnitude larger than for the optical region studied in
[\onlinecite{wang}].

%The large values of $\varepsilon_{\rm SC}$ means that the light almost does %not penetrate in the  cylinders. To determined the effect of the %penetration, we have studied a model structure, for which formally assumed %$\varepsilon_{\rm SC}\to\infty$, i.e. applied boundary conditions $E_z(\bm %\rho)=0$ at the cylinder surfaces. The spectra $R(\omega)$ and $T(\omega)$ %calculated for this structure look like the slightly blue-shifted spectra %$\alpha=1$. To explain this results

Below we propose a simple analytical description of photonic modes
near the edge of the allowed photonic band where the photon
frequency and two-dimensional Bloch wave vector ${\bm k} = (k_x,
k_y)$ are related by
\begin{equation} \label{dispersion}
\omega = \omega_0 +  C k^2\:.
\end{equation}
Here the zero-$k$ frequency $\omega_0$ and the coefficient
$\Omega$ have imaginary contributions if ${\rm Im} \{
\varepsilon_{\rm GK}(\omega) \} \neq 0$. Taking into account that
at low frequencies real part of $\varepsilon_{\rm GK}(\omega)$ is
negative and its modulus is very large, we can expand $\omega_0$
and $C$ in powers of $1/{\sqrt{- \varepsilon_{\rm GK}}}$. In the
first-order approximation we obtain instead of
Eq.~(\ref{dispersion})
\begin{equation}\label{fitw}
\omega =\omega_0^{\infty} - \frac{\Omega}{\sqrt{ -
\varepsilon_{\rm GK} (\omega_0^{\infty})}} + \left( \eta +
\frac{\zeta}{\sqrt{ - \varepsilon_{\rm GK}
(\omega_0^{\infty})}}\right)(ck)^2\:,
\end{equation}
where $c$ is the light velocity in vacuum (introduced here for
convenience), $\omega_0^{\infty}$ and $c^2 \eta$ are the values of
$\omega_0$ and $C$ in the limit $\varepsilon_{\rm GK} \to \infty$. It
should be stressed that the parameters $\omega_0^{\infty}$,
$\Omega$, $\eta$ and $\zeta$ depend only on the geometry of the
system and are independent of the cylinder dielectric function
$\varepsilon_{\rm SC}$. More precisely, one can use the
representation
\[
\omega_0^{\infty} = s_1\ \frac{2\pi c}{a}\:,\quad \Omega = s_2 \
\frac{2\pi c}{a}\:,\quad \eta = s_3\ \frac{a}{2\pi c}\:,\quad
\zeta = s_4\ \frac{a}{2\pi c}\:,
\]
where the coefficients $s_1$ to $s_4$ depend only on the ratio
$R/a$. The fitting of the first-allowed-band dispersion curves
obtained for different values of $\varepsilon_{\rm SC}$ give the
following values
\[
s_1 = 0.415\:,\:s_2 = 0.63 \:,\: s_3 = 0.98\:,\: s_4 = 3.45\:.
\]
Equation (\ref{fitw}) can be compared to the corresponding
expansion for a planar waveguide near the cutoff frequency, see [\onlinecite{sigalas}]. For an
ideal planar waveguide of the thickness $d$, one has
\[
\omega = c \sqrt{ \left( \frac{\pi}{d} \right)^2 + k^2 } \approx
\frac{\pi c}{d} + \frac{d}{2 \pi c}\ (ck)^2\:.
\]
The cutoff frequency and coefficient $\eta$ are equal to $\pi c/d$
and $d/2 \pi c$, respectively. On the other hand, for the PC under
study, the frequency $\omega_0^{\infty}$ is close to $\pi c/a$
(or, equivalently, $s_1 \approx 0.5$) and $s_3 \approx 1$. This
helps the interpretation of Eq.~(\ref{fitw}), namely, in the PC
the effective waveguide ``boundaries'' are the planes containing
cylinders. When the waveguide is not ideal, i.e., the dielectric
constant is negative, large but finite, the radiation slightly
penetrates in its walls which leads to an increase in the
effective thickness $d$ and a decrease in the cutoff frequency, in
agreement with the positive sign of $\Omega$.
\begin{figure}[t]
  \includegraphics[width=0.48\textwidth]{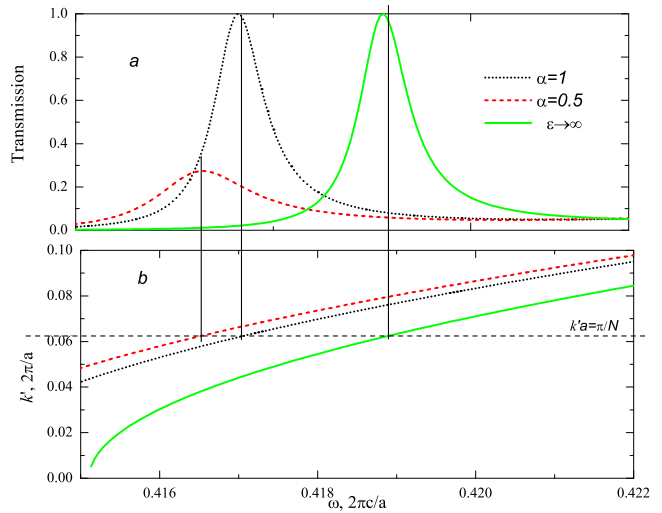}\\
  \caption{(a) Transmission spectrum $T(\omega)$ and (b) photon
dispersion $k'(\omega)$ near the first transmission peak. Solid,
dashed and dotted curves are calculated for $\alpha=1$,
$\alpha=0.5$, and $|\varepsilon_{\rm SC}| \to \infty$. Other
parameters including $N=8$ are the same as in Fig.~2. The vertical
lines indicate the frequencies at which $k'(\omega)=\pi/(Na)$,
this value of $k'$ is indicated by a horizontal line. }
\end{figure}

The relation between the transmission peak position and the photon
dispersion is analyzed in Fig.~3. Figure~3a presents the first
transmission peak for the PC structures with $\alpha = 0.5$,
$\alpha = 1$ and $\varepsilon_{\rm SC} \to \infty$. The first two
lines are fragments of those shown in Fig.~2a. The third line
corresponding to $\varepsilon_{\rm SC} \to \infty$ (equivalent to
the boundary condition $E_z(\bm \rho)=0$ at the cylinder surfaces)
looks like a slightly blue-shifted spectrum of the purely SC PC. A
value of the shift is in agreement with the estimation $\Omega
\left[ -\varepsilon_{\rm GK}( \omega_0^{\infty} ; \gamma = 0)
\right]^{-1/2}$ that follows from Eq.~(\ref{fitw}). This equation
also allows one to understand the evolution of the spectrum with
decreasing $\alpha$: for $\alpha$ different from unity the photon
wave vector acquires an imaginary part and the transmission
decreases. The similar analysis clearly shows that the spectral
position of the transmission maxima in Fig.~2 are determined by
the Fabry-P\'erot interference condition
\begin{equation}\label{fp}
N k'(\omega_m) a = m \pi,\quad m = 1,2 \ldots N-1\:,
\end{equation}
where $k'(\omega_m)$ is the real part of the photon wave vector at
the frequency $\omega_m$. Thus, the peaks in the transmission
spectrum can be related to the Fabry-P\'erot interference.
Extending this one-dimensional interpretation we can apply the
approximate estimation for the transmission and reflection
coefficients
\begin{equation} \label{T_N}
\begin{split}
T_N(\omega)= \frac{e^{- 2 k''(\omega)a N }|1 - r^2|^2}{|1 - r^2
{\rm e}^{2 {\rm i} k (\omega) a N }|^2},\\ R_N(\omega)= \frac{|r (1
- {\rm e}^{2 {\rm i} k (\omega) a N })|^2}{|1 - r^2 {\rm e}^{2
{\rm i} k (\omega) a N }|^2}
\end{split}
\end{equation}
derived for a slab of the thickness $Na$. Here $k''$ is the
imaginary part of the wave vector $k(\omega)$ calculated for the
photonic crystal and $r$ is the internal reflection coefficient
considered as a fitting parameter while making a comparison with
the exact calculation. Equation (\ref{fp}) implies the condition
$r''=0$ and, for lowest peak with $m=1$, the fit gives $r \approx
0.84$.

The two-dimensional nature of the PC makes it difficult to derive
a stricter analytical equations for the spectra, but the key
features of their dependence on $\alpha$, in particular, the red
shift and decrease of the absolute value of the transmission are
quite well reproduced by Eqs.~(\ref{T_N}) with $k(\omega)$
determined from Eq.~(\ref{fitw}). Fig.~4 presents the same curves
as Fig.~2 but in large scale and only in the vicinity of the peak
$m=1$.
 One can see that dashed curves, calculated by the
approximate equations (\ref{T_N}) lie sufficiently close to the
exact ones. Expressions (\ref{T_N}) can be simplified in the
spectral region near each frequency $\omega_m$ satisfying
Eq.~(\ref{fp}). Assuming $1 -r \ll 1$ and $k'' a N \ll 1 $, we can
reduce Eqs.~(\ref{T_N})
\begin{gather} \label{resa}
T_N(\omega)\approx \frac{\Gamma^{(0)2}_m}{(\omega-\omega_m)^2+
(\Gamma^{(0)}_{m}+\Gamma_m)^2 },\\  R_N(\omega)\approx
\frac{\Gamma_m^2+(\omega-\omega_m)^2}{(\omega-\omega_m)^2+
(\Gamma^{(0)}_{m}+\Gamma_m)^2 }\:, \nonumber\\
A_N(\omega)\approx
\frac{2\Gamma_m\Gamma_m^{(0)}}{(\omega-\omega_m)^2+
(\Gamma^{(0)}_{m}+\Gamma_m)^2 }\:, \nonumber
\end{gather}
with
\begin{equation}\label{Gamma}
\begin{split}
\Gamma_m = k''(\omega_m)aN \Delta,\quad \Gamma_m^{(0)}
=(1-r)\Delta,\\
\Delta=\frac{1}{aN} \left[\frac{dk'(\omega_m)}{d\omega}\right]^{-1}\:.\end{split}
\end{equation}
It follows then that narrow peaks in the transmission and
absorption as well as narrow dips in the reflection can be
described in terms of resonant tunneling in the region of a
``confined"-photon eigenfrequency $\omega_m$. Consequently,
$\Gamma_m^{(0)}$ and $\Gamma_m$ mean the radiative and
non-radiative decay rates of the corresponding ``confined'' state
$m$. The non-radiative decay vanishes for $\alpha=1$ and increases
with decreasing $\alpha$ which leads to a broadening of the peaks.
The peaks near the middle of the allowed band are wider than those
near the edge, as clearly seen in Fig.~2c, which is due to the
larger values of $\Delta \propto d \omega/dk' $. Note that,
although the integral over the absorption spectrum \eqref{resa}
monotonously increases with $\Gamma_m (\alpha)$, the peak value of
$A_N(\omega)$ has a maximum at $\Gamma_m(\alpha) =
\Gamma^{(0)}_m$, i.e., $ A_N(\omega_m, \Gamma_m = \Gamma_m^{(0)})
= 1/2$. Analogous behavior of the absorption peaks was observed
for the two-dimensional photonic crystal consisting of spherical
voids buried in a metal film~[\onlinecite{popov}]. If we take into account
overlapping between peaks (\ref{resa}) with different $m$ the
maximum value of the absorbance can slightly exceed 1/2, as is
seen in Fig.~2c.
\begin{figure}[t]
  \includegraphics[width=0.48\textwidth]{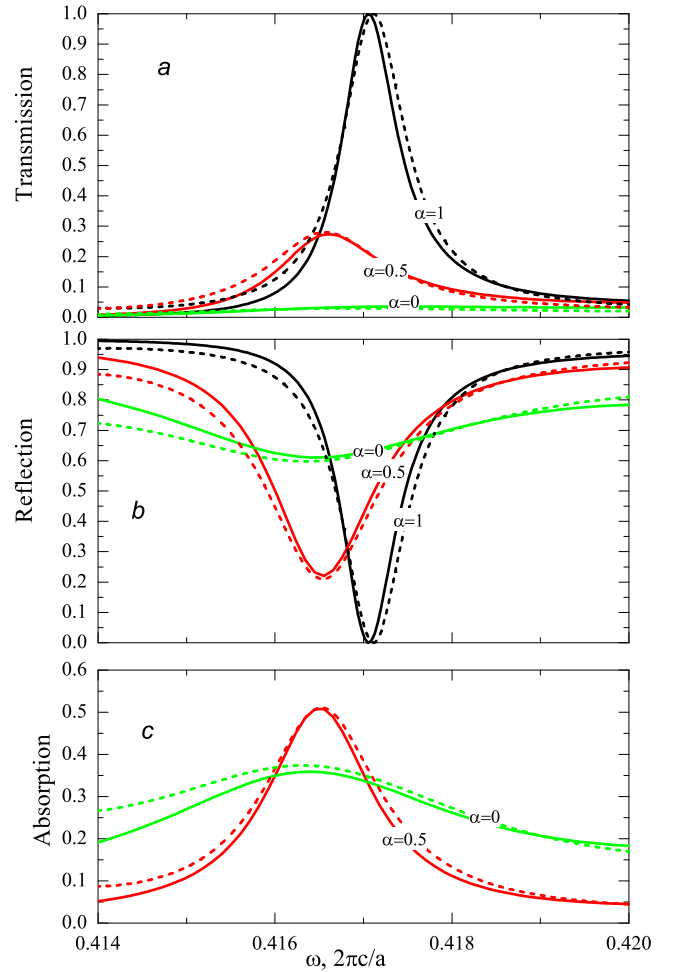}\\
  \caption{Transmission (a), reflection (b), and absorption (c) spectra of
a superconducting photonic crystal calculated for three indicated
values of $\alpha$ and the same set of parameters as in Fig.~2.
Solid and dashed curves correspond to spectra calculated
numerically and by using Eqs.~(\ref{T_N}). At $\alpha = 1$ the
absorption is zero, the corresponding line $A(\omega) \equiv 0$
coincides with the abscissa and is indistinguishable.
  }
\end{figure}

\section{Conclusion}
We have studied the photon dispersion law and optical properties
of a two-dimensional superconducting photonic crystal formed by
metallic cylinders arranged in a square lattice. The transmission,
reflection and absorption spectra are calculated as a function of
the fraction $\alpha$ of superconducting electrons. In the
spectral region within the allowed band the spectra exhibit
oscillations due to the interference effects. For the pure SC
phase, $\alpha = 1$, the absorption vanishes, $R(\omega) = 1
-T(\omega)$, and the transmission peak values reach unity. With
the transition from superconducting to normal metallic phase when
$\alpha$ changes from 1 to 0 the optical spectra exhibit
remarkable modifications. The proposed approximate analytical
description allows one to explain the key features of the
dispersion and spectral modifications with changing $\alpha$. The
peaks in transmission and absorption spectra as well as the dips
in the reflection spectra can be interpreted in terms of the
resonance tunneling.

\acknowledgements
%Author appreciates the valuable discussions with
%Profs. N.S. Averkiev, L.E. Golub, R.T. Harley, and E.L. Ivchenko.
The work was supported by RFBR  and the ``Dynasty'' foundation -- ICFPM.

%\bibliography{c:/work/coherent/bibliography/all}

\end{document}